\documentclass[11pt]{article}

\usepackage{fullpage}
\usepackage{epic,eepic,epsfig,graphics,amsmath,color,fullpage}
\usepackage{latexsym}
\usepackage[hang]{subfigure}
\usepackage{times}

{\renewcommand{\baselinestretch}{1}

\begin{document}


\newcommand{\cardi}[1]{ \vert #1 \vert}

\def\vval{{{\#}_{v}}}
\def\estval{{{\#}_{est_i}}}
\def\botval{{{\#}_{\bot}}}

\newtheorem{definition}{Definition}
\newtheorem{theorem}{Theorem}
\newtheorem{notation}{Notation}
\newtheorem{lemma}{Lemma}
\newtheorem{corollary}{Corollary}
\newcommand{\toto}{xxx}
\newenvironment{proofT}{\noindent{\bf
Proof }} {\hspace*{\fill}$\Box_{Theorem~\ref{\toto}}$\par\vspace{3mm}}
\newenvironment{proofL}{\noindent{\bf
Proof }} {\hspace*{\fill}$\Box_{Lemma~\ref{\toto}}$\par\vspace{3mm}}
\newenvironment{proofC}{\noindent{\bf
Proof }} {\hspace*{\fill}$\Box_{Corollary~\ref{\toto}}$\par\vspace{3mm}}

\newenvironment{theorem-repeat}[1]{\begin{trivlist}
\item[\hspace{\labelsep}{\bf\noindent Theorem~\ref{#1} }]}%
{\end{trivlist}}

\newenvironment{lemma-repeat}[1]{\begin{trivlist}
\item[\hspace{\labelsep}{\bf\noindent Lemma~\ref{#1} }]}%
{\end{trivlist}}

\newcounter{linecounter}
\newcommand{\linenumbering}{(\arabic{linecounter})}
\renewcommand{\line}[1]{\refstepcounter{linecounter}
\label{#1}
\linenumbering}
\newcommand{\resetline}{\setcounter{linecounter}{0}}

\newcommand{\grumbler}[2]{\begin{quote}{\sl \bf #1:} #2\end{quote}}


\title{\bf Time-Free and Timer-Based Assumptions Can
Be Combined to Solve Authenticated Byzantine Consensus}

\author{
 Hamouma Moumen   \\
 D\'epartement d'informatique,\\
  LMA, Universit\'e de Bejaia,
 06000 Bejaia , Algeria \\
 {\small{\tt hamouma.moumen@univ-bejaia.dz}}
}

\date{}
\maketitle
\renewcommand{\baselinestretch}{1,0}

\begin{abstract}
To circumvent the FLP impossibility result in a deterministic way
several protocols have been proposed on top of an asynchronous
distributed system enriched with additional assumptions. In the
context of Byzantine failures for systems where at most $t$
processes may exhibit a Byzantine behavior, two approaches have
been investigated to solve the consensus problem.The first, relies
on the addition of synchrony, called "Timer-Based", but the second
is based on the pattern of the messages that are exchanged, called
"Time-Free". This paper shows  that both types of assumptions are
not antagonist and  can be combined to solve authenticated
Byzantine consensus. This combined assumption considers a correct
process $p_i$, called $\diamond 2t$-BW, and a set $X$ of $2t$
processes such that, eventually, for each query broadcasted by a
correct process $p_j$ of $X$, $p_j$ receives a response from $p_i
\in X$ among the $(n - t)$ first responses to that query or both
links connecting $p_i$ and $p_j$ are timely. Based on this
combination, a simple hybrid authenticated Byzantine consensus
protocol,benefiting from the best of both worlds, is proposed.
Whereas many hybrid protocols have been designed for the consensus
problem in the crash model, this is, to our knowledge, the first
hybrid deterministic solution to the Byzantine consensus problem.

~
\end{abstract}







\section{Introduction}
\label{intro}
 \subsection{Context of the Study and Motivation}
 \label{context}
 The Consensus problem is one of the most attractive problems in the
the field  of asynchronous distributed systems. It may be used as
building block  to design or to implement several applications on
on top of fault prone asynchronous distributed systems, since it
abstracts several basic agreement problems. Solving the Consensus
problem in an  asynchronous distributed system where processes
(even only one) may crash is impossible\cite{FLP85}. This
impossibility result comes from the fact that it is impossible to
distinguish a crashed process from a process that is slow or with
which communication is  slow. To overcome this impossibility,
asynchronous distributed systems have to be enriched with
additional power such as Synchrony assumptions \cite{DLS88},
Common coins \cite{R83}, randomization \cite{B83, MMR14} ,
unreliable failure detectors \cite{CT96},and input vector
restrictions \cite{MRR03}. When considering the Consensus problem
in a setting where some processes can behave arbitrarily
(Byzantine behavior), solving this problem  becomes more
beneficial for designers or for developers of applications on top
of Byzantine fault prone asynchronous distributed system, but the
capacity of a such behavior, make this task more complex and more
difficult comparatively with crash failures. This difficulty comes
from the fact that a Byzantine process propose a wrong value and
it tries to impose it on correct processes.

\subsection{Related Works}
To solve the consensus with deterministic way , in the context of
crash failures, synchrony assumptions must be added \cite{DLS88}
or information about failures must be provided by a failures
detectors associated with the processes of the system \cite{CT96}.
A failure detector can be seen as a black box that gives (possibly
incorrect) information about process failures. Three  approaches
have been investigated to implement failures detectors. The first,
called "Timer-Based", considers the partially synchronous system
model \cite{CT96}, which is generalizes the model of \cite{DLS96},
where there are bounds on the relative speed of processes and
message transfer delays, but these bounds are not known and hold
only after some finite but unknown time, called Global
Stabilization Time (GST). The second approach, introduced in
\cite{MMR03}, does not assume timing assumptions about process
speeds and communication delays. This approach, called
"Time-Free", is based on the pattern of messages that are
exchanged. It considers the query-response-based winning messages
proposed in \cite{MMR03, MRT06} and the teta-model proposed in
\cite{WS09}. In the third approach,called hybrid ,  assumptions of
both approaches cited above are combined to implement failure
detectors \cite{MPR04}, \cite{MMRT06}.

In the context of Byzantine failure, the most of solutions for the
consensus problem consider the partially synchronous system model
where all links are eventually timely
\cite{ADFT06,CL99,DGG02,DGV05,KMM97,K02,L06,MA05}. In a such
context, the notion of failure detector, originally designed for
crash failures, is extended  to mute failures \cite{DGG02,KMM97}.
A muteness failure detector provides information about processes
that are silent ( did not send some consensus protocol messages).
This category is used directly in \cite{FMR05a,FMR05b} to solve
Byzantine consensus.

For the classical partially synchronous models \cite{CT96, DLS88}
composed of $n$ partially synchronous processes \cite{DLS88} among
where at most $t$ may crash, many models, that  require only some
links  which have to be timely, have been proposed
\cite{ADFT04,HMSZ06,MOZ05}in contrast of the related works cited
above which assume that the whole system is eventually
synchronous. The system model considered in \cite{ADFT04} assumes
at least an eventual $t$-source. An eventual $t$-source is a
correct process with $t$ outgoing eventually timely links
(processes communicate using point-to-point communication
primitives). On the other hand, the system model considered in
\cite{MOZ05} assumes a broadcast communication primitive and at
least one correct process with $t$ bidirectional but moving
eventually timely links. These two models are not comparable
\cite{HMSZ06}. In such a context, \cite{ADFT04} proved that an
$t$-source (eventual $t$-source) is necessary and sufficient to
solve consensus which means that it is not possible to solve
consensus if the number of eventually timely links is smaller than
$t$ or if they are not outgoing links of a same correct process.

For the second approach \cite{MMR03}, used to implement the
failure detectors defined in \cite{CT96}, where the are no
eventual bounds on process speeds and communication delays
(Message Pattern), \cite{MMRT06} proposed a leader protocol with
very weak assumption on the patten of messages that are exchanged.
This protocol assumes a correct process $p_i$ and a set $Q$,
possibly contains crashed processes, of $t$ processes (with $p \ni
Q$) such that, each time a process $p_j \in Q$ broadcasts a query,
it receives a response from $p_i$ among the first $(n-t)$
corresponding responses (such a response is called a winning
response). The two previous approaches ($\diamond t$-source and
Message Pattern) are combined , in \cite{MRT06}, to obtain an
eventual leader protocol. This combined assumption considers a
star communication structure involving $(t +1)$ processes (these
$t+ 1$ processes can differ from a run of the system to another
run) and is such that each of its $t$ links can satisfy a property
independently of the property satisfied by the $t - 1$ other
links.

In the context of Byzantine consensus where $t$ processes can
exhibit an arbitrary behavior, Aguilera et al. \cite{ADFT04}
propose a system model with weak synchrony properties that allows
solving the consensus problem. The model assumes at least an
$\diamond$bisource (eventual bisource). An $\diamond$bisource is a
correct process with all its outgoing and incoming links
eventually timely. This means that the number of eventually timely
links could be as low as $2(n-1)$ links. Their protocol does not
need authentication and consists of a series of rounds each made
up of 12 communication steps and $\Omega(n^3)$ messages. In
\cite{MMT07}  Moumen et al.  proposed a system model that
considers an eventual bisource with a scope of $2t$. The eventual
bisource assumed by \cite{ADFT04} has the maximal scope $(x = n .
1)$. An eventual $2t$-bisource ($\diamond 2t$-bisource) is a
correct process where the number of privileged neighbors is $2t$
where $t$ is the maximum number of faulty processes. Their
protocol needs authentication and consists of a series of rounds
each made up of 5 communication steps and $\Omega(n^2)$ messages.
In \cite{MM10}, Moumen and Mostefaoui  propose a weak system model
that does not rely on physical time but on the pattern of messages
that are exchanged. This model is based on the query-response
mechanism and assumes at least an $\diamond 2t$-winning process
(eventual $2t$winning process). An $\diamond 2t$-winning is is a
correct process where the number of privileged neighbors is $2t$ ,
such that eventually, for each query broadcasted by any of its
privileged neighbors , any of its privileged neighbors receives a
response from the $\diamond 2t$-winning process among the $(n -
t)$ first responses to that query. Their protocol needs
authentication and consists of a series of rounds each made up of
5 communication steps and $\Omega(n^2)$ messages. Note that this
assumption does not prevent message delays from always increasing
without bound. Hence, it is incomparable with the timer-based
$\diamond 2t$-bisource assumption.
\subsection{Contribution of the Paper}
The two previous approaches (Timer-Based and Time-free) have been
considered both in the case of crash failures and Byzantine
failures , but they have never been combined in the case of
Byzantine faults. This paper shows that timer-based and Time-Free
assumptions can be combined and proposes a system model where
processes are partially synchronous and the communication model
satisfies the requirements of the combined assumption. This
combined assumption consider a correct processes $p_i $, called
$\diamond 2t$-BW (B for Bisource and W for Winning), and a set $X$
of $2t$ processes (some processes may be Byzantine), such that
,eventually, for each query broadcasted by a correct process $p_j$
of $X$, $p_j$ receives a response from $p_i \notin X$ among the
$(n - t)$ first responses to that query or both links connecting
$p_i$ and $p_j$ are timely. In the case one all links that connect
$p_i$ with processes of $X$ then the $p_i$ is a $\diamond
2t$-bisource, but in the case one  all processes of $X$ receives
the response of $p_i$ among the $(n-t)$ response for each query
that have broadcasted, then $p_i$ is $\diamond 2t$-winning
process.

For the assumed model, a simple hybrid authenticated Byzantine
consensus protocol, benefiting from the best of both worlds, is
proposed. To our knowledge, this is the first protocol that
combines between Timer-Based and Time-Free Assumptions to solve
authenticated Byzantine consensus.
\subsection{Organization of the Paper}
The paper is made up of six sections. Section 2 presents the basic
computation model and the Consensus problem. Then, Section 3
 presents the  consensus protocol, with a $\diamond
2t$-BW, we propose and Section 4 proves its correctness.Finally,
Section 5 concludes the paper.

\section{Basic Computation Model and Consensus Problem}
\label{model}


\subsection{Asynchronous Distributed System with Byzantine Process }
We consider a message-passing system consisting of a finite set
$\Pi$ of $n$ $(n > 1)$ processes, namely,
$\Pi=\{p_1,\ldots,p_{n}\}$. A process executes steps (send a
message, receive a message or execute local computation). Value
$t$ denotes the maximum number of processes that can exhibit a
Byzantine behavior. A Byzantine process may behave in an arbitrary
manner. It can crash, fail to send or receive messages, send
arbitrary messages, start in an arbitrary state, send different
values to different processes, perform arbitrary state
transitions, etc. A correct process is one that does not
Byzantine. A faulty process is the one that is not correct.

Processes communicate and synchronize with each other by sending
and receiving messages over a network. The link from process p to
process $q$ is denoted $p \rightarrow q$. Every pair of process is
connected by two links $p \rightarrow q$ and $q \rightarrow p$.
Links are assumed to be reliable: they do not create, alter,
duplicate or lose messages. There is no assumption about the
relative speed of processes or message transfer delays.

\subsection{An authentication mechanism}
In order to deal with the power of Byzantine processes, We assume
that an authentication mechanism is available. A public key
cryptography such as RSA signatures \cite{RSA78} is used by a
process to verify the original sender of the message and to force a
process to relay the original message received. In our authenticated
Byzantine model, we assume that Byzantine processes are not able to
subvert the cryptographic primitives. To prevent a Byzantine process
to send different values to different processes, each message has to
carry a value and the set of $(n -t)$ values received by a process
during the previous step. The included signed values can be used by
a receiving process to check whether the value sent by any process
complies with the values received at the previous step. This set of
signed values is called certificate and its role is to prove to the
receiver that the value is legal.

To ensure the message validity, each process has an underlying
daemon that filters the messages it receives. For example, the
daemon will discard all duplicate messages (necessarily sent by
Byzantine processes as we assume reliable send and receive
operations between correct processes). The daemon, will also discard
all messages that are not syntactically correct, or that do not
comply with the text of the protocol.

\subsection{A Time-Free Assumption}
\paragraph{Query-Response Mechanism}
In this paper, we consider that each process is provided with a
query-response mechanism. More specifically, any process $p$ can
broadcast a QUERY () message and then wait for corresponding
RESPONSE () messages from $(n - t)$ processes. Each of this
RESPONSE () messages is a winning response for that query, and the
corresponding sender processes are the winning processes for that
query. The others responses received after the $(n - t)$ RESPONSE
() messages are the losing responses for that query, and
automatically discarded.  A process issues a new query only when
the previous one has terminated (the first $(n - t)$ responses
received). Finally, the response from a process to its
 own queries is assumed to always arrive among the first $(n - t)$ responses that is waiting for.

Henceforth, we reuse the definition of
\cite{MMR03,MPR04,MMRT06,MRT06} to define formally a winning
 link, an $x$- winning.

\begin{definition}
     Let $p$ and $q$ be two processes. The link $p \rightarrow q$
     is eventually winning (denoted $\diamond WL$) if there is a
     time $\tau$ such that the response from $p$ to each query issued
     by $q$ after $\tau$ is a winning response ($\tau$ is finite but unknown).
\end{definition}

\begin{definition}
 A process $p$ is an $x$-winning at time $\tau$ if $p$ is correct and there exists a set $X$ of processes of size
$x$, such that: for any process $q$ in $X$, the link $p
\rightarrow q$ is winning. The processes of $X$ are said to be
privileged neighbors of $p$.
\end{definition}

\begin{definition}
 A process $p$ is an $\diamond x$-winning if there is a time $\tau$ such that, for all $\tau' \geq \tau$, $p$ is an
$x$-winning at $\tau'$.
\end{definition}

\subsection{A Timer-Based Assumption}

Hereafter, we rephrase the definition of  \cite{HMSZ06} to define
formally a timely link and  an $x$- bisource.

\begin{definition} A link from a process $p_i$ to any process $p_j$ is timely
at time $\tau$ if (1) no message sent by $p_i$ at time $\tau$ is
received at $p_j$ after time ($\tau+\delta$) or (2) process $p_j$
is not correct.
\end{definition}

\begin{definition} A process $p_i$ is an $x$-bisource at time $\tau$ if:\\
- (1) $p_i$ is correct \\
- (2) There exists a set $X$ of processes of size $x$, such that:
for any process $p_j$ in $X$, both links from $p_i$ to $p_j$ and
from $p_j$ to $p_i$ are timely at time $\tau$. The processes of
$X$ are said to be privileged neighbors of $p_i$.
\end{definition}

\begin{definition} A process $p_i$ is an $\diamond x$-bisource if there is a time
$\tau$ such that, for all $\tau' \geq \tau$, $p_i$ is an
$x$-bisource at $\tau'$.
\end{definition}

\subsection{Combining Time-Free and Timer-Based Assumptions}

\begin{definition} A process $p_i$ is an $\diamond x$-BW at time $\tau$ if:\\
- (1) There exists a set $Y$ of processes of size $y$  and a set $Z$ of processes of size $z$ such that, $Y \cap Z = \emptyset$ and $y + z= x$\\
- (2) There is a time $\tau$ such that, for all $\tau' \geq \tau$,
      $p_i$ is an $y$-bisource and an $z$-winning at the same time
      $\tau'$. If $y=0$ then $p_i$ is an $x$-winning and if If $z=0$ then $p_i$ is an
      $x$-bisource.
\end{definition}

For the rest of the paper, we consider an asynchronous distributed
system where the only additional assumptions are those needed by
the $\diamond x$-BW.

\subsection{The Consensus Problem}
We consider the multivalued consensus problem, where there is no
bound on the cardinality of the set of proposable values. In the
multivalued consensus problem, every process $p_i$ proposes a value
$v$ and all correct processes have to eventually decide on a single
value among the values proposed by the processes.

Formally, the consensus problem is defined by the following three
properties:

 Let us observe that, in a Byzantine failure
context, the consensus definition should not be too strong. For
example, it is not possible to force a faulty process to decide the
same value as the correct processes, since a Byzantine process can
decide whatever it wants. Similarly, it is not reasonable to decide
any proposed value since a faulty process can initially propose
different values to distinct processes and consequently the notion
of ``proposed value'' may not be defined for Byzantine processes.
Thus, in such a context, the consensus problem is defined by the
following three properties:

\begin{itemize}
\item {\sf Termination}: Every correct process eventually decides.
\item {\sf Agreement}: No two correct processes decide different values.
\item {\sf Validity}: If all the correct processes propose the same
value $v$, then only the value $v$ can be decided.
\end{itemize}

\section{An Authenticated Byzantine Consensus Protocol With $\diamond 2t$-BW}
\label{hprot}
\begin{figure}
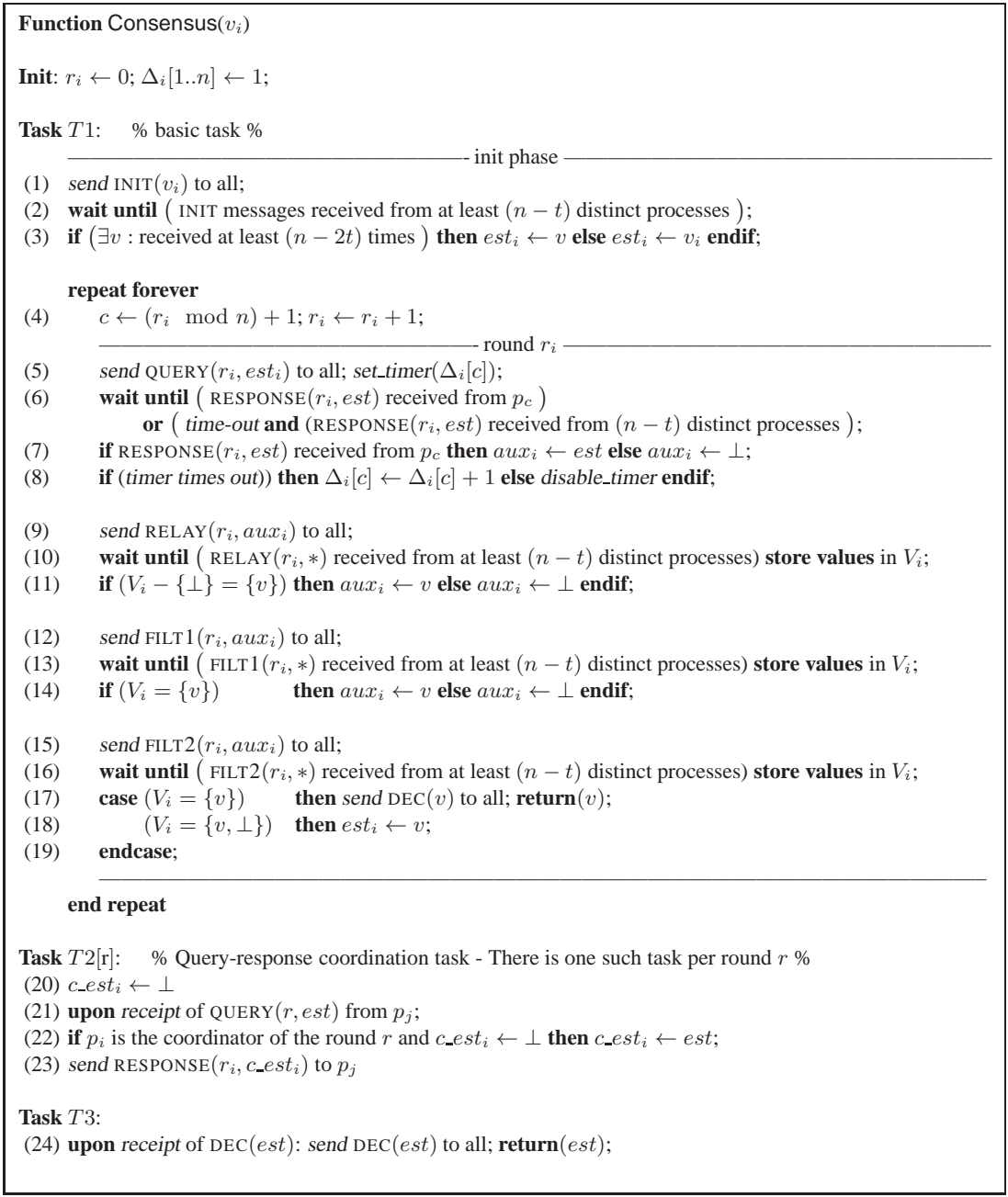

\centering{ \fbox{
\begin{minipage}[t]{150mm}
\footnotesize
\resetline
\begin{tabbing}
{\bf Function} {\sf Consensus}($v_i$)\\~\\

{\bf Init}: $r_i\leftarrow 0$; $\Delta_i[1..n] \leftarrow 1 $;\\~\\

{\bf Task} \= $T1$: ~~~ \% basic task \%\\

\> ------------------------------------------------------- init phase ----------------------------------------------------------- \\

\line{01} \> {\sl send} {\sc init}$(v_i)$ to all; \\

\line{02} \> {\bf wait until} \= \big(
 {\sc init} messages received from at least $(n-t)$ distinct processes \big); \\

\line{03} \> {\bf if} $\big( \exists v: $ received at least $(n-2t)$ times $\big)$ {\bf then} $est_i\leftarrow v$ {\bf else} $est_i\leftarrow v_i$ {\bf endif};\\~\\

\> {\bf rep}\={\bf eat forever}\\

\line{04} \>\> $c \leftarrow (r_i \mod n)+1$; $r_i \leftarrow r_i +1$; \\

\>\> ---------------------------------------------------- round $r_i$ ----------------------------------------------------------- \\

\line{05} \>\> {\sl send} {\sc query}$(r_i,est_i)$ to all; {\sl set\_timer}$(\Delta_i[c])$; \\

\line{06} \>\> {\bf wait until} \big( {\sc response}$(r_i,est)$  received from $p_c$ \big) \\

 \>\>  ~~~ ~~~  {\bf or} \big( {\sl time-out} {\bf and}  ({\sc
 response}$(r_i,est)$ received from $(n-t)$ distinct processes \big);\\

\line{07} \>\> {\bf if} {\sc response}$(r_i,est)$ received from
$p_c$ {\bf then} $aux_i \leftarrow est$ {\bf else} $aux_i
\leftarrow \bot$; \\

 \line{08} \>\> {\bf if} ({\sl timer times out}))
 {\bf then} $\Delta_i[c] \leftarrow \Delta_i[c]+1$
 {\bf else} {\sl disable\_timer} {\bf endif}; \\~\\


\line{09} \>\> {\sl send} {\sc relay}$(r_i,aux_i)$ to all; \\

\line{10} \>\> {\bf wait until} \= \big(
 {\sc relay}$(r_i,*)$ received from at least $(n-t)$ distinct processes) {\bf store values} in $V_i$; \\

\line{11} \>\> {\bf if} $(V_i-\{\bot\}=\{v\})$ \= {\bf then} $aux_i\leftarrow v$ {\bf else} $aux_i\leftarrow \bot$ {\bf endif};\\~\\


\line{12} \>\> {\sl send} {\sc filt1}$(r_i,aux_i)$ to all; \\

\line{13} \>\> {\bf wait until} \big(
 {\sc filt1}$(r_i,*)$ received from at least $(n-t)$ distinct processes) {\bf store values} in $V_i$; \\

\line{14} \>\> {\bf if} $(V_i=\{v\})$ \> {\bf then} $aux_i \leftarrow v$ {\bf else} $aux_i \leftarrow \bot$ {\bf endif};\\~\\


\line{15} \>\> {\sl send} {\sc filt2}$(r_i,aux_i)$ to all; \\

\line{16} \>\> {\bf wait until} \= \big(
 {\sc filt2}$(r_i,*)$ received from at least $(n-t)$ distinct processes) {\bf store values} in $V_i$; \\

\line{17} \>\> {\bf case} \=
 $(V_i=\{v\})$ ~~~~~~~ \= {\bf then} {\sl send} {\sc dec}$(v)$ to all; {\bf return}$(v)$; \\
\line{18} \>\>\> $(V_i=\{v,\bot\})$ \> {\bf then} $est_i \leftarrow v$; \\
\line{19} \>\> {\bf endcase};\\

\>\> -------------------------------------------------------------------------------------------------------------------------- \\

\> {\bf end repeat}\\~\\

{\bf Task} \= $T2[$r$]$: ~~~ \% Query-response coordination task - There is one such task per round $r$ \% \\
\line{20}\> $c\_est_i \leftarrow \bot$\\
\line{21}\> {\bf upon} {\sl receipt} of {\sc query}$(r,est)$ from $p_j$;\\
\line{22}\> {\bf if} $p_i$ is the coordinator of the round $r$ and $c\_est_i \leftarrow \bot$ {\bf then} $c\_est_i \leftarrow est$; \\
\line{23}\> {\sl send} {\sc response}$(r_i,c\_est_i)$  to $p_j$\\~\\
{\bf Task} \= $T3$: \\

\line{24}\> {\bf upon} {\sl receipt} of {\sc dec}$(est)$: {\sl send} {\sc dec}$(est)$ to all; {\bf return}$(est)$; \\

\end{tabbing}
\normalsize
\end{minipage}
} \caption{An Authenticated  Byzantine Consensus Protocol With
$\diamond 2t$-BW } \label{prot}}
\end{figure}

Figure \ref{prot} presents an authenticated Byzantine consensus
protocol in asynchronous distributed system where the only
additional assumptions are those needed by the $\diamond x$-BW.
The principle of the proposed protocol is similar to those that
have been proposed in \cite{MM10,MMT07} except the coordination
phase at the beginning of each round. Each process $p_i$ executes
the code of the protocol given by Figure \ref{prot}. This protocol
is composed of three tasks : a main task $(T1)$, a coordination
task $(T2)$, and a decision task $(T3)$.

 before executing the first round $(r=1)$,each process $p_i$ keeps its estimate of the decision value in
 a local variable $est_i$ and starts by the init phase in order to guarantee the
validity property.In this phase, each process $p_i$  sends {\sc
init}$(v_i)$ message , that containing its estimate, to all
processes.If $p_i$ receives at least $(n-2t)$ {\sc init} messages
for $v$ then it change its estimate to $v$, else it keeps its own
estimate. After this phase, the protocol proceeds in consecutive
asynchronous rounds. Each round $r$ is  composed of four
communication phases and is coordinated by a predetermined process
$p_c$ (line \ref{04}).

\noindent {\sl First phase of a round $r$}
(lines~\ref{05}-\ref{08}). Each process that starts a round
((including the coordinator of the round) first sends its own
estimate (with the associated certificate) to the coordinator
($p_c$) of the current round and sets a timer to ($\Delta_i[c]$).

In a separate task $T2[$r$]$(line \ref{21}), Each time a process
receives a valid QUERY message (perhaps from itself) containing an
estimate est, it sends a RESPONSE message to the sender. If the
process that responds to a query message is the coordinator of the
round to which is associated the query message, the value it sends
in the RESPONSE message is the coordination value. If the process
that responds is not the coordinator, it responds with any value
as the role of such a message is only to define winning links. as
the reader can find it in lines \ref{22}-\ref{23}, the value sent
by the coordinator is the value contained in the first valid query
message of the round it coordinates.

In the main task at line \ref{06}, a process $p_i$ waits for the
response from $p_c$ (the coordinator of the round) or for
expiration of the timer of $p_c$ ( $\Delta_i[c]$) and for $(n -
t)$ responses from others processes. In the latter case, process
$p_i$ is sure that $p_c$ is not the $\diamond 2t$-BW as its
response is not winning and its link with $p_i$ is not timely. If
a process $p_i$ receives a response from the coordinator then it
keeps the value in a variable $aux_i$ otherwise it sets $aux_i$ to
a default value $\bot$(this value cannot be proposed). If the
timer times out while waiting for the response from $p_c$,
$\Delta_i[c]$ is incremented and $p_i$ considers that its link
with $p_c$ is not eventually timely  or $p_c$ is Byzantine or the
value $\Delta_i[j]$ is not set to the right value. As
$\Delta_i[c]$ is incremented each time $p_c$'s responses misses
the deadline, it will eventually reach the bound on the round trip
between $p_i$ and $p_c$ if the link between between them is
timely.

If the current coordinator is a $\diamond 2t$-BW  then at least
$(t+1)$ correct processes will get the value $v$ of the
coordinator and thus set their variable $aux$ to $v$ ($\neq
\bot$). The next phases will serve to propagate this value from
the $(t+1)$ correct processes to all correct processes. Indeed,
among the $2t$ privileged neighbors of the current coordinator at
least $t$ are correct processes and all of them will receive the
value of the coordinator . If the current coordinator is
Byzantine, it can send nothing to some processes and/or perhaps
send different certified values to different processes . If the
current coordinator is not a $\diamond 2t$-BW or if it is
Byzantine, the three next phases allow correct processes to behave
in a consistent way. Either none of them decides or if some of
them decides a value $v$ despite the Byzantine behavior of the
coordinator, then the only certified value for the next round will
be $v$ preventing Byzantine processes from introducing other
values. The aim of the first phase is that if the coordinator is
an $\diamond 2t$-BW then at least
(t+1) correct process will get its value at the end of line \ref{07}.\\

\noindent {\sl Second phase of a round $r$}
(lines~\ref{09}-\ref{11}). During the second phase, all correct
processes relay, at line \ref{09}, either the value they received
from the coordinator (with its certificate) or the default value
$\bot$ if  they timed out and they received $(n - t)$ RESPONSE
messages from others processes. Each process collects $(n-t)$
valid messages and stores the values in a set $V_i$ (line
\ref{09}). At line \ref{09}, if the coordinator is correct only
one value is valid and can be relayed.

Moreover , if the current coordinator is a $\diamond 2t$-BW then
any correct process $p_i$ will get in its set $V_i$ at least one
copy of the value of the coordinator as among the $(t + 1)$ copies
sent by the $(t + 1)$ correct processes that got the value of the
coordinator a correct process cannot miss more than $t$ copies
(recall that a correct process collect $(n-t)$ valid messages). If
the coordinator is not a $\diamond 2t$-BW  or if
  it is Byzantine, some processes can receive only $\bot$ values, others may receive more
   than one value (the coordinator is necessarily Byzantine in this case) and some others
    can receive a unique value. This phase has no particular effect in such a case.
    The condition $(V_i-\{\bot\}=\{v\})$ of line \ref{11} means that if there is
    only one non-$\bot$ value $v$ in $V_i$ then this value is kept in $aux_i$ otherwise, $aux_i$ is set to $\bot$.
    The aim of this second phase is
that if the coordinator is an $\diamond 2t$-BW  then all
the correct processes will get its value.\\

\noindent {\sl Third phase of a round $r$}
(lines~\ref{12}-\ref{14}).This phase is a filter; it ensures that
at the end of this phase, at most one non\_$\bot$ value can be
kept in the $aux$ variables in the situations where the
coordinator is Byzantine. If the coordinator is correct, this is
already the case. When the coordinator is Byzantine two different
correct processes may have set their $aux_i$ variables to
different values. This phase consists of an all-to-all message
exchange.
   Each process collects $(n-t)$ valid messages the values of which are stored in a set $V_i$.
    If all received messages contain the same value $v$ ($V_i=\{v\}$) then $v$ is kept in $aux_i$
    otherwise $aux_i$ is set to the default value $\bot$. At the end of this phase, there is at most
    one (or none) certified value $v$ ($\neq \bot$).\\

\noindent {\sl Fourth phase of a round $r$}
(lines~\ref{15}-\ref{19}).

This phase ensures that the Agreement property will never be
violated. This prevention is done in the following way. If a
correct process $p_i$ decides $v$ during this round then if some
processes progress to the next round, then $v$ is the only
certified value. In this    decision phase, a process  $p_i$
collects $(n-t)$ valid messages and store the values in $V_i$. If
the set $V_i$ of a process $p_i$ contains a unique non\_$\bot$
value $v$, $p_i$ decides $v$. Indeed among the $(n-t)$ same values
$v$ received by $p_i$, at least $n-2t$ have been sent by correct
processes. As $(n-t)+(n-2t)>n$ any set of $(n-t)$ valid signed
messages of this phase includes at least one value $v$. Hence, all
processes receive at least one value $v$ (the other values could
be $v$ or $\bot$) and the only certified value for the next rounds
is $v$. This means that during the next round (if any) no
coordinator (whether correct or Byzantine) can send a valid value
different from $v$.

If during the fourth phase, a process $p_i$ receives only $\bot$
values, it is sure that no process can decide during this round
and thus it can keep the value it has already stored in $est_i$
(the certificate composed of the $(n-t)$ valid signed messages
received during phase four containing $\bot$ values, allow $p_i$
to keep its previous values $est_i$).

Before deciding (line \ref{17}), a process first sends to all
other processes a signed message {\sc dec} that contains the
decision value (and the associated certificate). This will prevent
the processes that progress to the next round from blocking
because some correct processes have already decided and stopped
sending messages. When a process $p_i$ receives a valid {\sc dec}
message at line \ref{24}, it first relays is to all other
processes and then decides. Indeed, task $T_3$ is used to
implement a reliable broadcast to disseminate the eventual
decision value preventing some correct processes from blocking
while others decide.


\section{Correctness of the protocol }
\label{proof}

\begin{lemma}\label{NABCL2}

 If tow corrects processes $p_i$ and $p_j$ decide $v$ and $v'$,
respectively, then $v = v'$.

\end{lemma}

\begin{proofL}
The proof is by contradiction. Suppose that $p_i$ and $p_j$ decide
$v$ and $v'$, respectively, such that $v \neq v'$. This means that
$v$ appears  at least $(n-t)$ times in $V_i$ and $v'$ also appears
at least $(n-t)$ times in $V_j$ at line \ref{16}. This means that
$\cardi{\mathcal{V}_i} + \cardi{\mathcal{V}_j}\geq 2(n-t)$. Since,
the $(n-t)$ correct processes send (according to the protocol) the
same message to both processes and the $t$ Byzantine processes can
send different messages to them, we have $\cardi{\mathcal{V}_i} +
\cardi{\mathcal{V}_j}\leq (n-t) + 2t = (n + t)$.  This leads to
$(n+t) \geq 2(n-t)$ i.e. $n\leq 3t$ a contradiction as we assume
$n>3t$.
\renewcommand{\toto}{NABCL2}
\end{proofL}
\begin{lemma}\label{NABCL3}
 If a correct process $p_i$ decides $v\neq \bot$ during a
round $r$ , then all correct processes start the next round with
the same estimate $v$ if they have not deciding.
\end{lemma}

\begin{proofL}
 Let us first note that if any correct process decides on
the value $v\neq \bot$ at the round $r$ then all correct
processes, that have not decided, set their estimates to $v$
because each of these processes receives at line \ref{22} at a
least one {\sc filt2}$(r_i,aux)$ message carrying the value $v$.
Moreover, all correct processes start a round $r+1$ with the same
estimate $v$.

The proof is by contradiction. Suppose that a correct process
$p_i$ decides $v$ at a round $r$ (line \ref{17}) and a correct
process $p_j$ has not decided at this round and sets own estimate
to $v'\neq v$. This means that the set $V_j$ of $p_j$ contains
only values different to $v$. By assumption, the value $v$,
appears in  $V_i$ at least $(n-t)$ times because it has decided.
As there are  $t$ Byzantine processes, $v$ is received by $p_i$ at
least $(n-2t)$ times  from correct processes. From these $(n-2t)$
messages for $p_i$ at most $t$ are loosed by $p_j$, because it
wait for $(n-t)$ messages at (line \ref{16}). From this, we can
conclude that $V_j$  contains at least $(n-3t)\geq$ times the
value $v$ ($n>3t$).
 Moreover, $p_j$ sets its estimate to $v$ . A contradiction.
\renewcommand{\toto}{NABCL3}
\end{proofL}

\begin{corollary}\label{NABCL4}
If a correct process $p_i$ decides a certified value $v$ during a
round $r$, then only $v$ can be decided in the same or in
subsequent rounds.
\end{corollary}

\begin{proofC}
Let us consider that a process $p_i$ decides a value $v$ in a
round $r$. If a correct process $p_j$ decides at the same round
$r$ then, by lemma  \ref{NABCL2}, it decides the same value $v$
decided by $p_i$. If a correct process $p_j$ does not decide at
the same round $r$ then, by lemma \ref{NABCL3}, all correct
processes start the next round $r+1$ by the same estimate value
$v$ decided by $p_i$ at a round $r$.Indeed, in the latter case,
$v$ will be the only certified value as even $\bot$ is not
certified as a certificate for the value that will be used during
the next round is composed by a set of $(n - t)$ messages as we
said above that any such set includes at least one value $v$. From
now on, the only value that can be exchanged is $v$ and only $v$
can be decided

\renewcommand{\toto}{NABCL4}
\end{proofC}
\begin{theorem}[agreement]
\label{WR1}

 No two correct
processes decide differently.
\end{theorem}

\begin{proofT}
If a correct process decides at line \ref{24}, it decides a
certified value decided by another process. Let us consider the
first round where a process decides at line \ref{17}. By Corollary
\ref{NBACL4},if a process decides a certified value during the
same round, it decides the same value. If a process decides after
receiving a {\sc dec} message at line \ref{24} it decides the same
value. Any process that starts the next round with its local
variable $est_i \neq v$ will see its messages rejected (no value
different from $v$ can be certified).
\renewcommand{\toto}{WR1}
\end{proofT}


\begin{lemma}\label{NABCL8}
If no process decides  during a round $ r' \leq r$, then all
correct processes start round $r + 1$.
\end{lemma}

\begin{proofL}
Let us first note that a correct process cannot be blocked forever
in the init phase. Moreover, it cannot be blocked at line \ref{06}
because of the time-out and  at least $(n - t)$ processes respond
to QUERY messages.

Suppose that no process has decided a value $v$ during a round $r'
\leq r$, where $r$ is the smallest round number in which a correct
process $p_i$ blocks forever. The proof is by contradiction.

By assumption, $p_i$ is blocked at lines \ref{10}, \ref{13} or
\ref{16}.

Let us first examine the case where $p_i$ blocks at line \ref{10},
which is the first statement of round $r$ where a process can
block forever. This means that at least $(n-t)$ correct processes
eventually execute line \ref{09}, because processes are partially
synchronous. Consequently as communication is reliable between
correct processes the messages sent  by correct processes will
eventually arrive at  $p_i$ that blocks forever at line \ref{10}.
The cases where $p_i$ blocks at line \ref{13}or \ref{16}  are
similar to this first case. It follows that if $p_i$ does not
decide, it will proceed to the next round. A contradiction.
\renewcommand{\toto}{NABCL8}
\end{proofL}

\

\begin{theorem}[termination]
\label{NABCL10}

 If there is a $\diamond 2t$-BW in the system, then all correct processes eventually
decide.
\end{theorem}

\begin{proofT}
As the protocol uses authentication, if some process receives a
valid DEC message, it can decide even if the message has been sent
by a faulty process. Recall that a Byzantine process cannot forge
a signature. If a correct process decides at line \ref{17} or at
line \ref{24} then, due to the sending of DEC messages at line
\ref{17} or line \ref{24}, respectively, prior to the decision,
any correct process will receive such a message and decide
accordingly (line \ref{24}).\\

So, suppose that no correct process decides. The proof is by
contradiction. By hypothesis, there is a time $\tau$ after which
there is a process $p_x$ that is a $\diamond 2t$-BW. Let $p_j$ be
a correct process and one of the $2t$ privileged neighbors of
$p_x$. Let $r$ be the first round that starts after $\tau$ and
that is coordinated by $p_x$. As by assumption no process decides,
due to Lemma \ref{NABCL8}.

 All correct processes $p_i$ (and possibly some Byzantine
processes) start round r and send a valid QUERY message to $p_x$
(line \ref{05}). This QUERY message contains a value $est$ which
is the estimate of process $p_i$. When the coordinator $p_x$ of
round $r$ receives the first QUERY message (line \ref{21})
possibly from itself, it sets a local variable $c\_est_x$ to the
valid value contained in the message. Then each time process $p_x$
receives a QUERY message related to this round $(r)$, it sends a
RESPONSE message to the sending process. If we consider any
correct process $p_i$ privileged neighbor of $p_x$, the RESPONSE
message from $p_x$ the coordinator to the QUERY message of $p_i$
will be received by $p_i$ among the first $n - t$ responses
because the link between $p_i$ and $p_x$ is winning or the
RESPONSE message from $p_x$ will be received by $p_i$ before
expiration of $\Delta_i[x]$ , because the link between $p_i$ and
$p_x$ is timely (line \ref{21}).

In the worst case, there are $t$ Byzantine processes among the
$2t+1$ privileged neighbors of $p_x$. A Byzantine process can
either relay the value of $p_x$ ($t$ Byzantine processes, $t$
correct processes and itself). During the next phase,a Byzantine
process can either relay the value of $p_x$ or relay $\bot$
arguing that $\Delta_i[x]$ has expired and it did not receive the
response of $p_x$ among the first $(n-t)$ RESPONSE messages (the
value of $p_x$ and $\bot$ are the only two valid values for this
round). This allows to conclude that the value $v$ sent by $p_x$
the coordinator of the present round is relayed at line \ref{09})
by, at least, the $t+1$ correct processes with which $p_x$ has
timely or winning links (the only other possible value is
$\bot$).

During the third phase (lines \ref{12}-\ref{14}), as the value $v$
of $p_x$ is the only certified value , all the processes that send
a certified message at line \ref{12} . This allows to conclude
that all processes will have to set their variables $aux$ variable
to $v$ (line \ref{14}). By the same way, all processes that send
certified messages at line \ref{15} will send $v$. From there we
can conclude that correct processes will all decide at line
\ref{17}, which proves the theorem.
\renewcommand{\toto}{NABCL10}
\end{proofT}

\begin{theorem}[Validity]
\label{sec:strongCons} If all correct processes propose $v$, then
only $v$ can be decided.
\end{theorem}

\begin{proofT}
Let $v$ be the only proposed value by correct processes at line
\ref{01}. Since all correct processes propose $v$, $v$ is sent at
least $(n-t)$ times at line \ref{01}. Since processes receive  at
least $(n-t)$ values from distinct processes, we can conclude that
at a line \ref{03} the values $v$ is received  at least $(n-2t)$
times by any correct processes.
  Moreover, any value proposed by Byzantine processes will be received at most $t$ times. As $n>3t$,
  we have $t<n-2t$. Consequently, the only certified value is $v$. This means that all correct
   processes set their variable $est$ to $v$.

\renewcommand{\toto}{sec:strongCons}
\end{proofT}

\section{Conclusion}
\label{conclusion}
This paper has shown that timer-based assumption and time-free
assumption can be combined to solve authenticated Byzantine
consensus in asynchronous distributed systems. It has presented
the first deterministic  authenticated Byzantine  protocol that
benefiting from the best of both worlds. This combined assumption
considers a correct process $p_i$, called $\diamond 2t$-BW, and a
set $X$ of $2t$ processes such that,eventually, for each query
broadcasted by a correct process $p_j$ of $X$, $p_j$ receives a
response from $p_i \in X$ among the $(n - t)$ first responses to
that query or both links connecting $p_i$ and $p_j$ are timely.
The proposed protocol has very simple design principle and it
provides an assumption coverage better than the one offered by any
protocol based on a single of these assumptions. In favorable
setting, the proposed protocol can reach decision in only 6
communication steps and needs only $\Omega(n^2)$ messages in each
step. The major contribution of this paper is to show that
Byzantine Consensus is possible with a very weak hybrid additional
that satisfying  the properties required by a $\diamond 2t$-BW .




\begin{thebibliography}{99}
{\small

\bibitem{ACT97}
Aguilera M.K., Chen W., and Toueg S., Heartbeat:
a timeout-free failure detector for quiescent
reliable communication. {\it Proc Workshop on Distributed
Algorithms (WDAG'97)}, pages 126-140, 1997.

\bibitem{ADFT04}
Aguilera M.K., Delporte-Gallet C., Fauconnier H. and Toueg S,
Communication-efficient leader election and consensus
with limited link synchrony.
{\it Proc. 23nd ACM Symposium on Principles of Distributed Computing (PODC'04)},
{\sc acm p}ress, 2004.

\bibitem{ADFT06}
Aguilera M.K., Delporte-Gallet C., Fauconnier H. and Toueg S., Consensus
with byzantine failures and little system synchrony.
{\it Proc. International Conference on Dependable Systems and Networks
(DSN'06)}, Philadelphia, 2006.

\bibitem{BCT96}
Basu A., Charron-Bost B., and Toueg T., Crash
failures vs. crash + link failures. {\it Proc 15th ACM Symposium on Principles
of Distributed Computing (PODC'96)}, Philadelphia,
Pennsylvania, 1996.

\bibitem{B83}
Ben-Or M.,
Another Advantage of Free Choice: Completely Asynchronous Agreement
Protocols.
{\it Proc. 2nd ACM Symposium on Principles of Distributed Computing
(PODC'83)}, {\sc acm p}ress, pp. 27-30, 1983.


\bibitem{CL99}
Castro, M. and Liskov, B., Practical Byzantine fault tolerance. {\it Proc.
of the 3rd Symposium on Operating Systems Design and Implementation},
New Orleans, USA, February 1999.

\bibitem{CT96}
Chandra T.D.~and Toueg S.,
Unreliable Failure Detectors for Reliable Distributed Systems.
{\em Journal of the ACM}, 43(2):225-267, 1996.

\bibitem{CNLV04}
Correia M., Neves N.F., Lung L.C. and Verissimo P., Low Complexity
Byzantine-Resilient Consensus. {\it Distributed Computing}, Volume
17, 13 pages, 2004.

\bibitem{DGG02}
Doudou A., Garbinato B. and Guerraoui R.,
Encapsulating Failure Detection: from Crash to Byzantine Failures.
{\it Proc. International Conference on Reliable Software Technologies},
Vienna (Austria), 2002.

\bibitem{DGV05}
Dutta P., Guerraoui R., and Vukolic M., Best-case complexity
of asynchronous byzantine consensus. {\it Technical Report
EPFL/IC/200499}, EPFL, Feb. 2005.

\bibitem{DLS88}
Dwork C., Lynch N.A. and Stockmeyer L., Consensus in
the presence of partial synchrony. {\em Journal of the ACM}, 35(2):288-323,
1988.

\bibitem{FLP85}
Fischer M.J., Lynch N. and Paterson M.S.,
Impossibility of Distributed Consensus with One Faulty Process.
{\em Journal of the ACM}, 32(2):374-382, 1985.

\bibitem{FMR05a}
Friedman R., Mostefaoui A. and Raynal M., Simple and efficient
oracle-based consensus protocols for asynchronous byzantine
systems. {\em IEEE Transactions on Dependable and Secure
Computing}, 2(1):46-56, 2005.

\bibitem{FMR05b}
Friedman R., Mostefaoui A., Raynal M., $\diamond$P-Mute-Based
Consensus for Asynchronous Byzantine Systems. Parallel Processing
Letters, vol 15(1-2):169-182, 2005.

\bibitem{HMSZ06}
Hutle M., Malkhi D., Schmid U., and Zhou L.,
Chasing the Weakest System Model for Implementing.
{\it Research Report 74/2005, Technische Universit\"at Wien,
Institut f\"ur Technische Informatik}, July, 2006.

\bibitem{K02}
Kursawe K., Optimistic Byzantine agreement. {\it Proc. of the 21st
IEEE Symposium on Reliable Distributed Systems (SRDS'02
Workshops)}, 2002.


\bibitem{KMM97}
Kihlstrom K.P., Moser L.E., and Melliar-Smith P.M.,
Solving Consensus in a Byzantine Environment Using an Unreliable Fault
Detector,
{\it Proc. of the 1st Int. Conference on Principles of Distributed Systems (OPODIS'97)}, pp. 61-75, 1997.



\bibitem{L06}
Lamport, L., Lower bounds for asynchronous consensus. {\it Distributed
Computing}, vol 19(2):104-125, 2006.

\bibitem{MA05} Martin J.P., and Alvisi L., Fast Byzantine paxos.
{\it Proc. International Conference on Dependable Systems and
Networks (DSN'05)}, Yokohama, Japan, pp.\,402-411, 2005.

\bibitem{MM10}
H. Moumen and A. Mostefaoui: Time-Free Authenticated Byzantine
Consensus {\it Proc . on the Ninth IEEE International Symposium on
Network Computing and Applications} pp.140-146, Cambridge, USA,
2010.
\bibitem{MMR14}
Most\'efaoui A., Moumen H., and Raynal M., Signature-free
asynchronous Byzantine consensus with $t<n/3$ and $O(n^2)$
messages. {\it Proc. 33th  ACM  Symposium on Principles of
Distributed Computing (PODC'14)}, ACM Press,  pp.~2-9, 2014.


\bibitem{MMR03}
 A. Mostefaoui, E. Mourgaya, and M. Raynal, "Asynchronous Implementation of Failure Detectors,
 {\it Proc. Int'l IEEE Conf. Dependable Systems and Networks} (DSN '03), pp. 351-360, 2003

\bibitem{MMRT06}
 Achour Mostefaoui, Eric Mourgaya, Michel Raynal, Corentin Travers: A Time-free Assumption to Implement Eventual Leadership.{\it Parallel Processing Letters} {\bf 16(2)}: 189-208 (2006)

\bibitem{MMT07}
 Hamouma Moumen, Achour Mostefaoui, Gilles Tredan: Byzantine Consensus with Few Synchronous Links.{\it Proc. of the Int. Conference on Principles of Distributed Systems} (OPODIS), pp. 76-89. 2007.

\bibitem{MOZ05} Malkhi D., Oprea F., and Zhou L.,
$\Omega$ meets paxos: Leader election and stability without
eventual timely links. {\it Proc. 19th International Conference on
Distributed Computing (DISC'05)}, Cracow, Poland, pp.\,26-29,
2005,
\bibitem{MPR04}
 Achour Mostefaoui, David Powell, Michel Raynal: A Hybrid Approach for Building Eventually Accurate Failure Detectors. {\it PRDC} 2004: 57-65

\bibitem{MRR03}  Mostefaoui A., Rajsbaum S, Raynal M, Conditions on Input Vectors for Consensus Solvability
in Asynchronous Distributed Systems. {\em Journal of the ACM}, 50
(6), pp. 922–954, November 2003.

\bibitem{MRT06} Achour Mostefaoui, Michel Raynal, Corentin Travers, "Time-Free and Timer-Based Assumptions Can Be Combined to Obtain Eventual Leadership,
{\it IEEE Transactions on Parallel and Distributed Systems}, vol. {\bf 17}, no. 7, pp. 656-666, Jul., 2006



\bibitem{PSL80}
Pease L., Shostak R. and Lamport L.,
Reaching Agreement in Presence of Faults.
{\em Journal of the ACM}, 27(2):228-234, 1980.

\bibitem{R83}
Rabin M.,
Randomized Byzantine Generals.
{\em Proc. 24th IEEE Symposium on Foundations of Computer Science
(FOCS'83)}, pp. 403-409, 1983.

\bibitem{RSA78}
Ronald L. Rivest, Adi Shamir, and Leonard M. Adleman. A method for
obtaining digital signatures and public-key cryptosystems. {\it
Communications of the ACM}, {\bf 21(2)}:120-126, February 1978.

\bibitem{S90}
Schneider F.B.,
Implementing Fault-Tolerant Services Using
the State Machine Approach: A Tutorial.
{\em ACM Computing Surveys}, 22(4):299-319, 1990.

\bibitem{ST87}
Srikanth T.K. and Toueg S., Simulating authenticated broadcasts to
derive simple fault-tolerant algorithms. {\it Distributed
Computing}, 2(2):380-394, 1987.
\bibitem{WS09} Josef Widder, Ulrich Schmid: The Theta-Model: achieving synchrony without clocks. {\it Distributed Computing}, {\bf 22(1)}: 29-47 (2009).


}
\end{thebibliography}
\end{document}